\documentclass[sort&compress,preprint]{elsarticle}
\usepackage{geometry}
\usepackage{setspace}
\usepackage{amsmath,graphicx} 
\usepackage{siunitx}
\usepackage{braket}
\usepackage{amssymb}
\usepackage{gensymb}
\usepackage{caption}
\usepackage{subcaption}
\usepackage{mathtools}
\usepackage[thinc]{esdiff}
\usepackage[capitalise]{cleveref}
\usepackage{xfrac}
\usepackage[dvipsnames]{xcolor}
\usepackage[normalem]{ulem}
\usepackage{multirow}
\usepackage{leftidx}
\usepackage{mathrsfs}

\usepackage{float}
\usepackage{siunitx}

\journal{Computational Material Science}

\begin{document}

\begin{frontmatter}
\onecolumn
\title{Development of a Model for Irradiation-Assisted Grain Growth for Nanocrystalline UO$_2$}
\author[INL_Fuel_Development]{Md Ali Muntaha}
\author[INL]{Larry K. Aagesen}
\author[UF]{Michael R. Tonks\corref{mycorrespondingauthor}}
\cortext[mycorrespondingauthor]{Corresponding author}
\ead{michael.tonks@ufl.edu}

\address[INL_Fuel_Development]{Fuel Development Performance and Qualification, Idaho National Laboratory, Idaho Falls, ID 83415, USA}
\address[INL]{Computational Mechanics and Materials, Idaho National Laboratory, Idaho Falls, ID 83415, USA}
\address[UF]{Department of Materials Science and Engineering, University of Florida, Gainesville, FL 32611, USA}

\begin{abstract}
In this work, we have developed a model for irradiation-assisted grain growth in nanocrystalline UO$_2$ using the MARMOT code. We include the impact of irradiation on UO$_2$ grain growth by coupling a phase field grain growth model with a heat conduction simulation that features a random heat source representing thermal spikes. Our model parameters have been calibrated against experimental measurements at 300 K. The calibrated model predicts grain growth in an irradiated UO$_2$ thin film that compares well with experimental data at 50 K. These results suggest that thermal spikes are the major cause of the irradiation-assisted grain growth observed in the UO$_2$ experiments. They also indicate that irradiation-assisted grain growth is only significant with average grain sizes less than 35 nm, and thus can be neglected when considering fuel performance of typical UO$_2$ fuel pellets.

\end{abstract}


\end{frontmatter}
\newpage

\section{Introduction}
\label{sec:introduction}
Grain growth occurs in polycrystalline materials to reduce their total free energy~\cite{Mullins_1956}. Various driving forces, such as a reduction in grain boundary (GB) energy, stored elastic energy, and stored defect energy, contribute to grain growth \cite{gottstein2009grain}. The reduction of GB energy decreases the total area of GBs, which have a higher energy compared to the interior of the grains. The process involves the growth of larger grains and the disappearance of smaller ones, causing a decrease in the number of grains and an increase in the average grain size. The kinetics of grain growth are typically defined by the GB mobility, which often has an Arrhenius dependence on the temperature. Thus, grain growth typically occurs only at elevated temperatures, especially in ceramics such as UO$_2$~\cite{Tonks-PC-Jake_2021,Tonks_letter_2014,MOELANS_2009_479,Nichols_2004}.

However, radiation can induce grain growth at much lower temperatures than is necessary for thermally-activated grain growth \cite{kaoumi2008thermal,KAOUMI_2006490,Zhang_2010,VOEGELI_2003_230,bufford2015unraveling, Zefeng_2022, Motta_2021, Kaoumi2008_conf}, including cryogenic temperatures. For example, Kaoumi et al. showed accelerated grain growth in ion-irradiated metallic thin films of Zr, Cu, Pt, and Au \cite{kaoumi2008thermal}, and supersaturated solid solutions of Zr–Fe and Cu–Fe \cite{KAOUMI_2006490}. Later, Yu et al.~\cite{Zefeng_2022} showed irradiation-accelerated grain growth for in-situ ion-irradiated UO$_2$ thin films. Kaoumi et al.~\cite{kaoumi2008thermal} developed an analytical model that predicted the evolution of the average grain size in irradiated materials based on the direct impact of thermal spikes on GBs, which accelerated curvature-driven growth. Bufford et al.~\cite{bufford2015unraveling} developed a mesoscale model of grain growth in which they introduced randomly distributed regions of locally increased GB mobility and compared the predictions with data from in situ ion irradiations of Au thin films. 

The grain size of UO$_2$ reactor fuel influences its performance by directly affecting the fission gas release, heat conduction, creep, and fracture \cite{TURNBULL_1974_62,Shrestha_2019,TOMA_FGR_2D,MUNTAHA_FGR_2D,GONG2019169,Pieterjan_2021,WATANABE_2008_388,DongUk-Kim,PASTORE_2013_75,PASTORE_2015_398,Oguma_1982, Chatterjee2025_fissiongas_UO2}. For this reason, a large number of experiments~\cite{AINSCOUGH_1973_117} and atomistic~\cite{Nerikar_2011,SUN_2014137} and mesoscale simulations~\cite{Tonks_letter_2014,AHMED_2014_90,AHMED_2017_25,GUO_2018_24,HALLBERG_2015_664,Tonks_2015_GB_Pinning} have been carried out to investigate GB migration and grain growth in UO$_2$, as summarized by Tonks et al.~\cite{Tonks-PC-Jake_2021}. However, these studies have focused on thermally-driven GB migration and have not considered irradiation-assisted grain growth. Therefore, there is a need to develop models that consider irradiation-assisted grain growth in UO$_2$ to determine if it has a significant impact on fuel performance. 

In this work, we have included irradiation-assisted grain growth in a phase field grain growth model for UO$_2$. We have compared the predictions of the model with the in situ grain growth data from the irradiation experiments carried out by Yu et al.~\cite{Zefeng_2022} in UO$_2$ thin films. We begin by summarizing the experimental data from Yu et al.\ in Section \ref{sec:Experimental Findings}.  We discuss the model formulation in Section \ref{sec:Model_development}, including the grain growth and heat conduction models, the numerical methods, and the domain size and parameter selection. In Section \ref{sec:results}, we present results and we discuss them in Section \ref{sec:discussion}. Finally, we conclude in Section \ref{sec:conclusions}.

\section{Experimental observation of irradiation assisted grain growth in UO$_2$}
\label{sec:Experimental Findings}

In the range of cryogenic to room temperatures, the migration of GBs under irradiation is generally assumed to be associated with the thermal spikes that take place during the process of radiation-induced damage. When an energetic particle strikes a target material, the kinetic energy of the particle is deposited into a target atom. This transfer of energy takes the target atom to a high-energy state, making it mobile and displacing it in the lattice. These displaced atoms can, in turn, collide with other atoms in the lattice, creating a chain reaction of collisions, which is called the collision cascade. When a high-energy ion, e.g.~an energy $\geq 5.42\times10^{12}$ eV/nm$^3$s~\cite{Motta_2021}, impacts a material surface, an extremely high-energy thermal spike event is created, which lasts for picoseconds~\cite{Motta_2021,Zefeng_2022}. When a thermal spike occurs at or near a GB, it incites an increase in the number of atomic displacements across the boundary. This action results in a movement of the GB, which ultimately leads to accelerated grain growth.

\begin{figure}[tbp]
\centering
\includegraphics[width=0.6\linewidth]{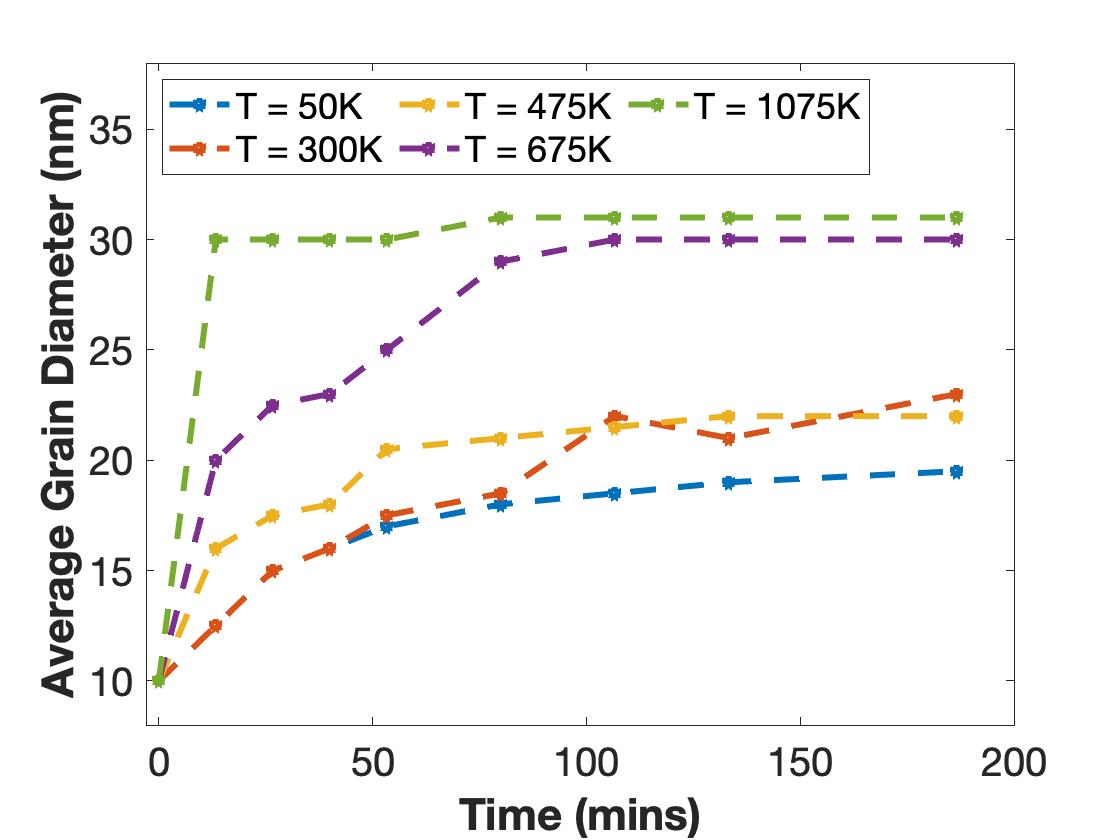}
\caption{The average grain diameter versus time from in situ ion irradiations of UO$_2$ nanocrystallien thin films carried out by Yu et al.~\cite{Zefeng_2022} at various temperatures. The time was computed by dividing the reported fluences by the irradiation flux.}
\label{fig:Experimental_Data_Temperatures}
\end{figure}

Yu et al.~\cite{Zefeng_2022} used in situ ion irradiation to investigate irradiation-assisted grain growth in UO$_2$. They performed isothermal irradiation experiments on UO$_2$ thin films (50 nm thickness) using $1$ MeV Kr$^{2+}$ ions at a flux of $6.25\times10^{15}$ ions/m$^2$s, reaching a maximum fluence of $7\times10^{19}$ ions/m$^2$. The irradiations were carried out for approximately 3 hours at temperatures ranging from 50 K to 1075 K. Yu et al. \cite{Zefeng_2022} measured the grain diameter using transmission electron microscopy (TEM) images that were collected by stopping the irradiations at regular intervals. Details on how the grain diameter was collected are available in their paper \cite{Zefeng_2022}. For all temperatures, grain growth occurred rapidly in the early stages and then slowed significantly over time, as shown in~\cref{fig:Experimental_Data_Temperatures}. The average thermal spike energy $q_c$ and the number of thermal spikes per ion $\chi$ were calculated using the Stopping and Ranging of Ions in Matter (SRIM) code \cite{STOLLER_201375}. 

To determine the temperatures at which thermally activated grain growth is significant in their irradiated samples, Yu et al.~\cite{Zefeng_2022} also carried out grain growth experiments at the same temperatures without irradiation. They observed no thermally activated grain growth after a 5-hour annealing period for temperatures up to 475 K. Thus, only irradiation-assisted grain growth was significant in 50 K, 300 K and 475 K
irradiations (see~\cref{fig:Experimental_Data_Temperatures}). Thermally-activated grain growth did occur above 475 K, indicating that both thermally-activated and irradiation-assisted grain growth were significant in the 675 K and 1075 K irradiations. Yu et al.~\cite{Zefeng_2022} also found that the thermal spike diameter was temperature dependent for $T > 300$ K, with larger diameters observed at higher bulk material temperatures. Due to these results, we focus on modeling the behavior at temperatures of 300 K and below, where we can assume that the spike diameter is independent of temperature and that thermally activated grain growth is negligible.

\section{Model development}
We have developed a mesoscale model of irradiation-assisted grain by coupling a phase field grain growth model with a heat conduction model that includes heat generation due to thermal spikes arising from collision cascade events. In this section, we summarize the model and include all the assumptions involved in its development. We begin by summarizing the phase field grain growth model and the heat conduction model. We then discuss its numerical implementation and the parameter selection.

\label{sec:Model_development}

\subsection{Phase field grain growth model}
\label{sec:Model_development_Phase_Field}

In phase field grain growth models, instead of tracking the precise location of sharp GBs, the phase field method represents the grain structure using smooth, continuous field variables $\eta_i$, called order parameters. A specific order parameter $\eta_i=1$ in grain $i$, $\eta_i=0$ in all other grains $j \neq i$, and $0<\eta_i<1$ across GBs between grain $i$ and other grains. The order parameters describing the grains evolve to minimize the free energy of the system  according to the Allen-Cahn equation
\begin{equation}
    \frac{\partial{\eta_i}}{\partial{t}}=-L \frac{\delta{F}}{\delta{\eta_i}},\ \mathrm{for}\ i = 1,..., N, \label{Eq:AC_evolution}\end{equation}
where $N$ is the total number of order parameters, $L$ is the order parameter mobility, $F$ is the free energy functional, and $\frac{\delta}{\delta \eta_i}$ represents the variational derivative with respect to $\eta_i$. We use the phase field model developed by Moelans et al.~\cite{moelans2008quantitative}, in which the free energy functional is
\begin{equation}
    F =\int_{V}\,  \left( \sum_i^N m\left( \frac{\eta_i^4}{4} - \frac{\eta_i^2}{2}+ \frac{3}{2} \sum_i^N\sum_{j>i}^N \eta_i^2 \eta_j^2 + \frac{1}{4} \right) +  \frac{\kappa}{2} \sum_i^N |\nabla \eta_i|^2 \right)dV, \label{eq:total_free_energy}
\end{equation}
where $m$ and $\kappa$ are model parameters that define the GB energy of the material and the width of the diffuse interfaces used to represent the GBs. Moelans et al.~\cite{moelans2008quantitative} developed quantitative expressions that define $m$, $\kappa$, and $L$ in terms of the interface width $l_{int}$, GB energy $\gamma$, and the GB mobility $M_{GB}$:
\begin{align}
    m\ &= 6 \frac{\gamma}{l_{int}} \\
    \kappa &= \frac{3}{4} \gamma l_{int} \\
    L &= \frac{4}{3} \frac{M_{GB}}{l_{int}}.
\end{align}

In our model, we assume all GB types have the same energy and mobility. Expressions for the average GB energy and mobility for UO$_2$ are presented by Tonks et al.~\cite{tonks2021mechanistic}. The average GB energy is a function of temperature according to
\begin{equation}
    \gamma = \left(1.56-5.87\times10^{-4}T\right)\ \mathrm{J/m^2}.
\end{equation}
The average GB mobility follows the standard Arrhenius expression
\begin{equation}
    M_{GB} = M^0_{GB} \mathrm{e}^{-\frac{E_a}{R T}},
\end{equation}
where $M_{GB}^0=2.14\times10^{-7}$ m$^4$/(J s) is the prefactor, $E_a=290$ kJ/mol is the activation energy, $R$ is the ideal gas constant constant, and $T$ is the temperature. We assume that the GB mobility is defined by this expression even for regions within the thermal spike that can locally reach above the melting temperature.



\subsection{Heat conduction model}
\label{sec:Model_development_Heat_Conduction}

To model irradiation-assisted grain growth, we have to modify the phase field grain growth model to include the impact of thermal spikes due to collision cascades on GB migration. The idea of a thermal spike in a material has been confirmed by analytical models and molecular dynamics simulations \cite{de1987role}. Bufford et al.~\cite{bufford2015unraveling} developed a phase field grain growth model that included the impact of thermal spikes by increasing the GB mobility in randomly selected regions representing the thermal spike. They showed that this approach resulted in predicted grain growth behavior that was qualitatively similar to that observed in radiation experiments. In our model, we couple the grain growth equations to a heat conduction equation that includes randomly distributed regions of short-lived heat generation, which elevate the local temperature, often above the melting point, for a short period of time. 

We solve for the change in the temperature field $T$ with time $t$ across the polycrystal by solving the heat equation:
\begin{equation}
    \rho c_p \frac{\partial T}{\partial t} = \nabla \cdot k \nabla T + Q, \label{eq:heat_equation}
\end{equation}
where $\rho$ is the density, $c_p$ is the specific heat, $k$ is 
the thermal conductivity, and $Q$ is the heat generation due to thermal spikes. 

The heat generation field $Q$ varies stochastically in space across the domain and in time due to the random nature of thermal spikes from cascades occurring in the system. A given rate of cascades per unit volume per unit time, $P_c$, was used to calculate the probability of a cascade occurring. It is a function of the radiation flux $\phi$ according to
\begin{equation}
    P_c = \chi \, \phi, \label{eq:probability}
\end{equation}
where $\chi$ is the number of cascades generated per ion per unit length. When a cascade is randomly determined to occur during a certain time step, the center location of the cascade is randomly determined from a uniform distribution. The value of $Q$ is then non-zero in a spherical region of radius $r_s$ around the center point for a specific thermal spike duration $t_s$ of the order of picoseconds. The magnitude of the heat generation rate within the spherical region is
\begin{equation}
    Q = \frac{q_s}{V_s t_s}, \label{eq:heat_gen}
\end{equation}
where $q_s$ is the average thermal spike energy in eV per spike and $V_s = 4/3\ \pi r_s^3$ is the spike volume. 

\begin{figure}[tbp]
\centering
\includegraphics[width=0.98\linewidth]{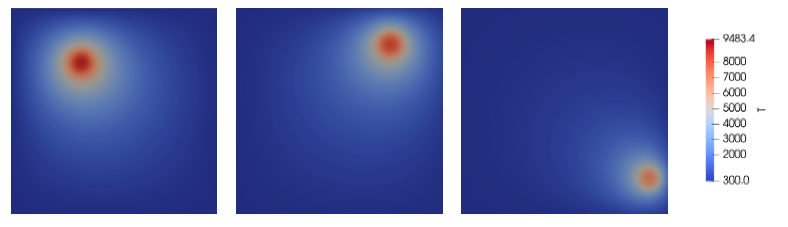}
\caption{Demonstration of the implementation of thermal spike events using the heat conduction model for a domain size of $100$ nm $\times$ $100$ nm. Temperature profiles are displayed at three different time steps, where thermal spike events can be observed at three different random locations, each with the peak temperatures at the center of the spikes.}
\label{fig:Temperature_profile_heat_conduction_only}
\end{figure}

The local heat generation due to thermal spikes results in local increases in temperature that can increase well above the melting temperature of UO$_2$, as demonstrated in ~\cref{fig:Temperature_profile_heat_conduction_only}. For this demonstration, the initial temperature is 300 K in a $100$ nm $\times$ $100$ nm domain. We assume Dirichlet boundary conditions of 300 K on all boundaries to represent heat transport from the sample boundaries that are in contact with the environment. The temperature rises as high as 9483.4K at the center of a thermal spike. After the spike duration, the heat quickly dissipates through the material and the local temperature returns to room temperature.

\subsection{Domain Size and Parameter Selection}
\label{sec:domain_size_parameter}

To simulate the ion irradiations from Yu et al.~\cite{Zefeng_2022}, we use 2D simulations with a square domain. We use 2D simulations rather than 3D to reduce the computational cost. We assume an initial average grain size of 10 nm, which is consistent with the average grain sizes from the experimental thin films. We assume columnar grains, such that the grain structure does not change through the thickness of the films. We also assume no heat transport from the top and bottom of the films and that the thermal spikes are cylinders rather than spheres. Together, these assumptions result in the grain growth not changing in the $z$-diretion, allowing us to model irradiation-assisted grain growth in 2D. We assume a domain thickness in the $z$-direction of $4/3 r_s$, such that volume of the thermal spike cylinder in our 2D domain is equivalent to the thermal spike sphere in a 3D domain.

Using a domain size as large as the actual thin films in the experiments would result in too many grains to be computationally feasible. Therefore, we select a smaller domain size to provide a more reasonable number of grains. We use a 710 nm by 710 nm domain with 5000 initial grains. The grain structure is constructed from a Voronoi tessellation and then thermally-activated grain growth is simulated to obtain a more realistic 5000 grain structure. 

\begin{figure}[tbp]
\centering
\includegraphics[width=0.6\linewidth]{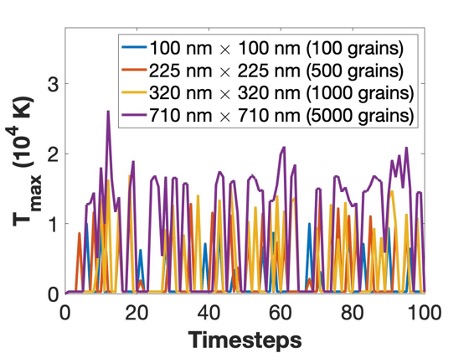}
\caption{Demonstration of the peak temperatures over time for different domain sizes. The legend also indicates the number of initial grains that can be accommodated in that domain with an average grain size of 10 nm.}
\label{fig:Temperature_peak_with_domain_size}
\end{figure}

Natural boundary conditions are used for the order parameters $\eta_i$. For the temperature, heat is removed from the system at the sample edges by the holder. Thus, we apply Dirichlet boundary conditions at all boundaries with a value of the irradiation temperature. However, due to the temperature boundary conditions, the peak temperatures that occur vary with the domain size, as shown in \cref{fig:Temperature_peak_with_domain_size}. The smaller the domain size, the more the boundary conditions reduce the maximum temperatures during a thermal spike. Thus, it is desirable to use as large a domain size as is computationally feasible. We find that the 710 nm by 710 nm square domain provides the best balance between computational expense and accuracy.

All of the parameters for our simulations are either taken from the literature for UO$_2$ or are chosen to represent the ion irradiation experiments from Yu et al.~\cite{Zefeng_2022}. All of the properties used in the model can be found in Table~\ref{table: Experimental and MD data for phase field simulations}, including the references where the values were obtained.  
 \begin{table}[btp]
\renewcommand\arraystretch{2.0}
\caption{The parameter values used in our irradiation-assisted grain growth model. The reference that is the source of the value is provided, where applicable.}
\begin{center}
\begin{tabular}{m{5cm} m{1.5 cm}  m{4cm}  m{1.5cm}}
\hline\hline 
Parameter  & Symbol & Value  &  Reference \\ 
\hline
Spike radius & $r_s$          &    4.84 nm              & \cite{kaoumi2008thermal, ULMER_2021_152688}\\
Spike duration &  $t_s$                           &    1-100 ps              & \cite{Stuchbery_1999,HEREDIAAVALOS_2021_152459} \\
Average thermal spike energy & $q_s$                          &   25.65 KeV             & \cite{kaoumi2008thermal, ULMER_2021_152688} \\
Thermal spike generated per ion per length &  $\chi$  &  0.0405 spikes/ion-nm   & \cite{kaoumi2008thermal,ULMER_2021_152688}\\
Ion flux          &       $\phi$           &  $6.25\times10^{11}$ ions/cm$^2$-s              & \cite{Motta_2021}\\
Initial average grain size &                 & 10 nm                         & \cite{Zefeng_2022,Motta_2021} \\
Incident ion energy/type    &        & 1 MeV Kr                        & \cite{Zefeng_2022,Motta_2021} \\
Irradiation fluence         &        & 0 to $7\times10^{19}$ ions/m$^2$   & \cite{Zefeng_2022,Motta_2021} \\
Interfacial width & $l_{int}$ & 2 nm & \\
Thermal conductivity & $k$             &  3 W/m-k                          & \cite{K_Baranov,K_GIBBY_1971163,K_GODFREY,K_Kim,K_Kingery,K_osti_Bates,K_Sheindlin,K_PhysRevLett.110.157401,K_WANG2015267,K_Ronchi,K_FINK20001} \\
Molar heat capacity & $C_o$          &  213.955 J/mol-k     &    \cite{K_Ronchi,K_FINK20001,Co_ENGEL1969211,Co_GOEL2008438,Co_GRONVOLD1970665,Co_HUNTZICKER197161,Co_INABA1987341,Co_TAKAHASHI1993108,Co_YUN2012109}       \\ 
UO$_2$ specific heat & $C_p$           &   0.7925 J/g-K       &\cite{Motta_2021} \\
GB mobility prefactor & $M_{o}$      &   2.14$\times$ $10^{-7}$ m$^4$/{J-s}   & \cite{Tonks-PC-Jake_2021} \\
Activation energy & $E_a$            &   3 eV                      & \cite{Tonks-PC-Jake_2021} \\
GB energy & $\sigma_{GB}$           &  Eq. (6)                         & \cite{tonks2021mechanistic} \\
UO$_2$ density & $\rho_{UO_2}$               &   10.97 g/cm$^3$       & \\
UO$_2$ molar mass &              &   0.27 kg/mol       & \\


\hline\hline
\end{tabular}
\end{center}

\label{table: Experimental and MD data for phase field simulations}
\end{table}

To represent the ion irradiation conditions from Yu et al.~\cite{Zefeng_2022}, we use the ion flux applied in their work, $\phi=6.25\times10^{12}$ ions/cm$^2$s. We use a thermal spike radius $r_s = 4.84$ nm and an average spike energy $q_c = 25.65$ keV, as described in the experimental work~\cite{Zefeng_2022}. The thermal spike duration $t_s$ is unknown, but potentially ranges from around one to hundreds of ps. We determine the spike duration by comparing with the experiments, as discussed in \cref{sec:results}. 

\subsection{Numerical Implementation}
\label{sec:numerical}
We implement our irradiation-assisted grain growth model using the finite element method in the mesoscale nuclear materials tool MARMOT \cite{tonks2012object}, which is based on the Multiphysics Object-Oriented Simulation Environment (MOOSE) framework \cite{permann2020moose}. We use implicit time integration with a second-order backward Euler scheme to simultaneously solve Eq.~\eqref{Eq:AC_evolution} to evolve each of the order parameters $\eta_i$ and Eq.~\eqref{eq:heat_equation} to evolve the temperature $T$. 

To reduce the computational cost of polycrystal grain growth simulations, each order parameter represents multiple grains. We use the GrainTracker algorithm in MOOSE \cite{PERMANN201618} to remap grains to new order parameters to avoid unphysical grain coalescence. We use 11 order parameters to represent all of the grains.

We solve the resultant system of nonlinear equations using the Preconditioned Jacobian Free Newton Krylov (PJFNK) method. Our nonlinear solution converges with a relative tolerance of 10$^{-8}$ and an absolute tolerance of 10$^{-10}$. We have utilized MOOSE's automatic scaling feature to address the substantial variation in the magnitude of the order parameter and temperature residuals. 

\begin{figure}[htbp]
\centering
\includegraphics [width=0.6\linewidth]{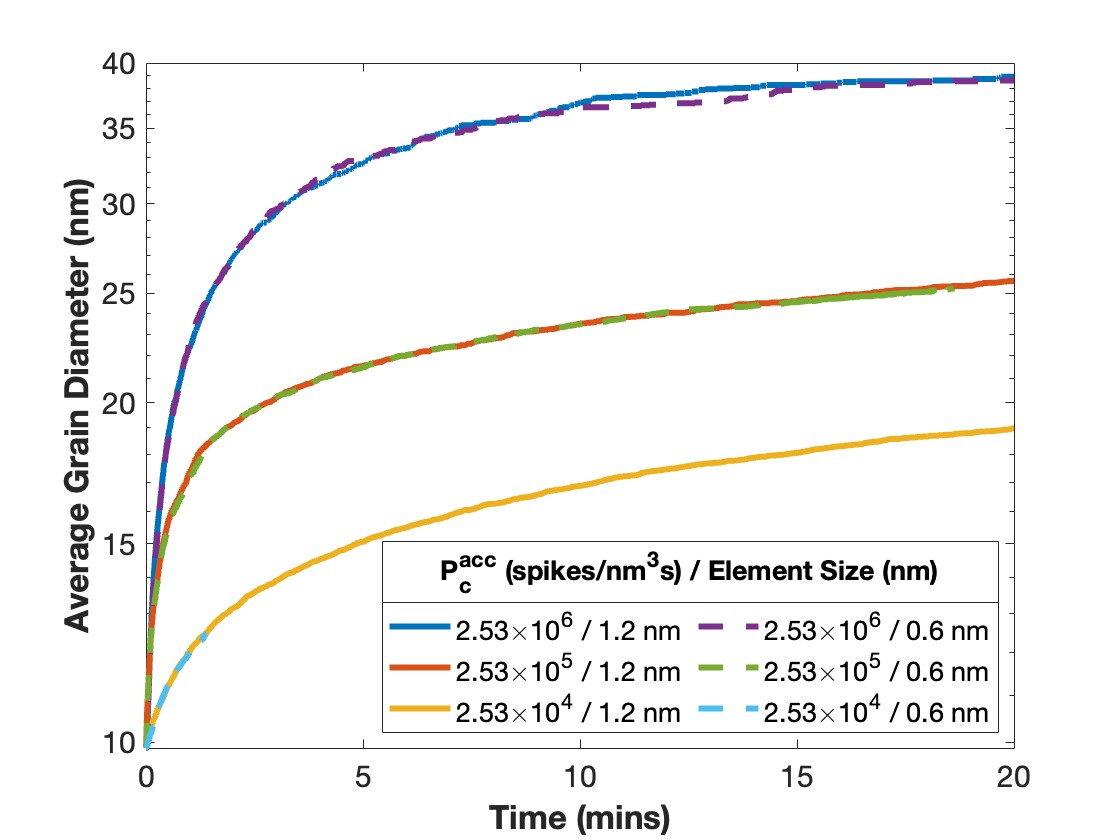}
\caption{Comparison of the predicted grain growth using an element size of 0.6 nm (shown with dashed lines) and 1.2 nm (shown with solid lines).}
\label{fig:mesh_refinement_effect}
\end{figure}

We use square elements to discretize the domain. The side length of the elements needs to be small enough to adequately resolve the interfacial width of 2 nm (see \cref{table: Experimental and MD data for phase field simulations}). We compare the predicted grain growth using element side lengths of 0.6 nm and 1.2 nm for a range of different spike rate cases, and find little difference in the predicted behavior as shown in in~\cref{fig:mesh_refinement_effect}. Therefore, we use the larger element size of 1.2 nm to reduce the computational cost.

\begin{figure}[htbp]
\centering
\includegraphics[width=0.6\linewidth]{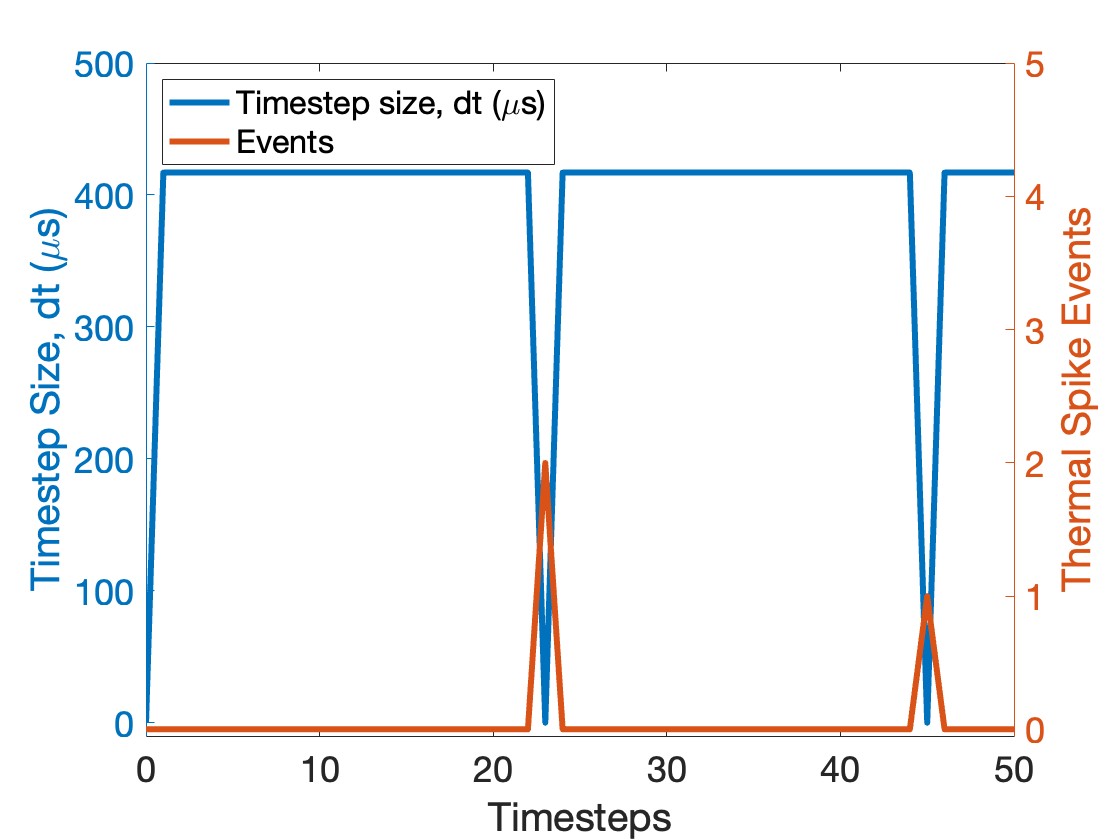}
\caption{Demonstration of the decrease in the time step during a thermal spike.}
\label{fig:dt_events_timestep}
\end{figure}

To further reduce the computational expense, we use adaptive time stepping. The time step adapts to maintain a target of 10 nonlinear iterations in the PJFNK solve per time step. When a thermal spike occurs, the time step automatically decreases to the spike duration $t_s$, to ensure that it is accurately represented. This connection between thermal spike events and the time step size is illustrated in~\cref{fig:dt_events_timestep}.

Even with adaptive time stepping, the requirement to resolve the spike duration makes the computational cost of the simulations unreasonably high. Assuming $\chi = 0.0405$ spikes per ion per nm \cite{kaoumi2008thermal,ULMER_2021_152688} and our ion flux, Eq.~\eqref{eq:probability} gives a value of $P_c=0.00253$ spikes/nm$^3$s. Given our 710 nm by 710 nm domain size with an assumed thickness of $4/3r_s$, this means on average a thermal spike will occur every 0.15 ms. Since our time step size goes down to the spike duration (1-100 ps) during each thermal spike there could be millions of time steps between each thermal spike. 
However,3 since the temperature is too low between thermal spikes to have any GB migration, the GBs only migrate during thermal spikes. Thus, we can reduce the computational cost by not modeling the time between thermal spikes and still accurately capture the irradiation-assisted grain growth.

To avoid modeling the time between thermal spikes, we increase the cascade rate $P_c$ to the point where only one or two thermal spikes occur every time step. We then scale the time by the ratio of the rates, where the actual time can be computed from the accelerated time:
\begin{align}
    \frac{P_c^{act}}{P_c^{acc}} &= \frac{t_{acc}}{t_{act}} \nonumber \\
    t_{act} &= \frac{P_c^{acc}}{P_c^{act}} t_{acc}, 
    \label{eq:P_acc}
\end{align}
where $P_c^{act}$ and $t_{act}$ are the actual rate and time and $P_c^{acc}$ and $t_{acc}$ are the accelerated rate and time. This approach is valid if only one spike occurs per time step or if multiple spikes occur but are far enough apart that they do not interact. 

\begin{figure}[tbp]
\centering
\begin{subfigure}{.49\textwidth}
  \centering
  \includegraphics[width=0.99\linewidth]{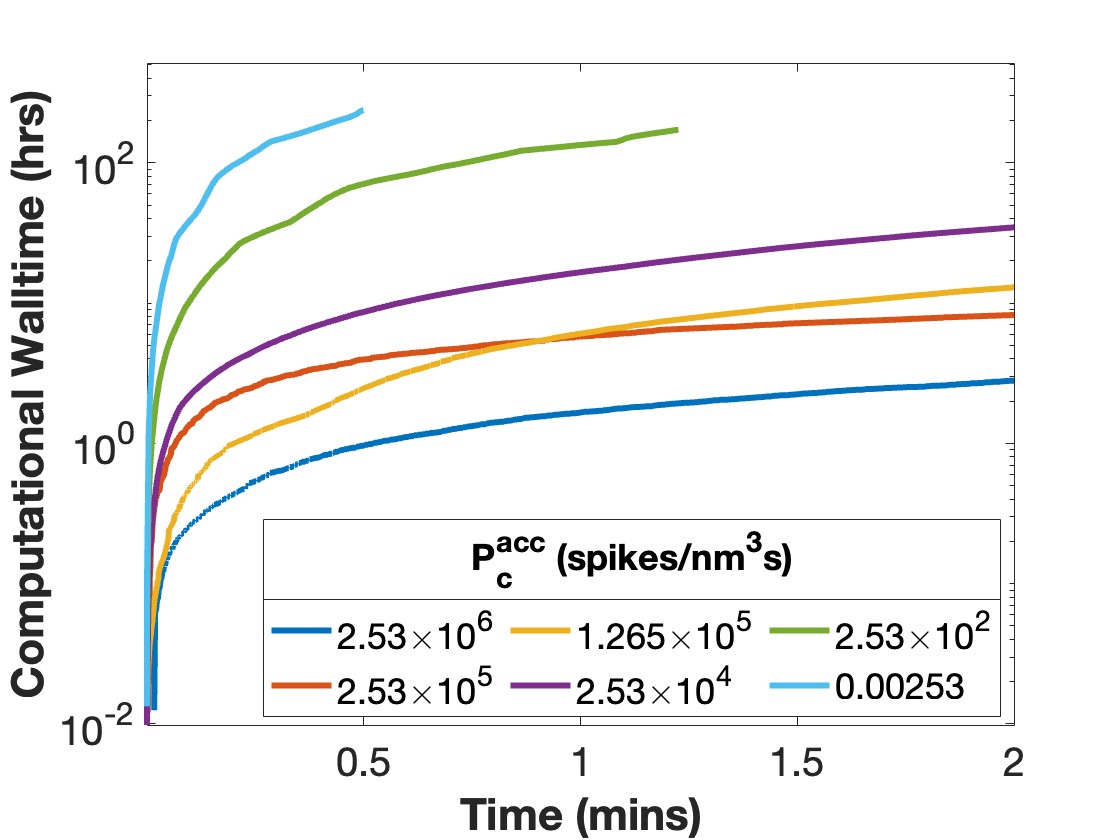}
  \caption{}
  \label{fig:Impact-of-probability-walltime}
\end{subfigure}%
\begin{subfigure}{.49\textwidth}
  \vspace{0.5cm}
  \centering
  \includegraphics[width=0.99\linewidth]{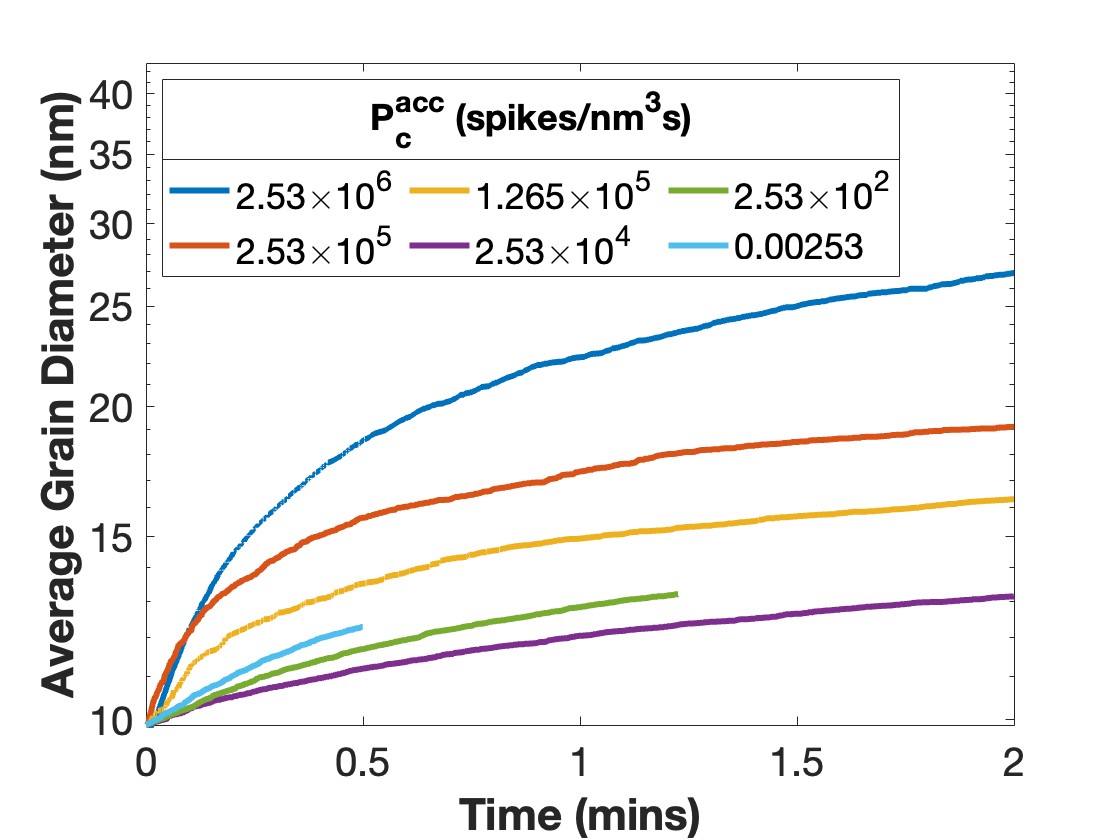} 
  \caption{}
  \label{fig:Impact-of-probability-grain-size}
\end{subfigure}%
\caption{Impact of $P_c^{acc}$ on (a) the wall time and (b) the average grain diameter with time. The results use a 710 nm by 710 nm domain with $t_s=100$ ps.}
\label{fig:Impact-of-probability}
\end{figure}

Preliminary results with different rate values and assuming a spike duration $t_s = 100$ ps are shown in~\cref{fig:Impact-of-probability}. Each simulation was carried out using 160 processors on the University of Florida HiperGator supercomputer. The computational walltime is excessively high when considering the actual rate of $0.00253$ spikes/nm$^3$s, as shown in \cref{fig:Impact-of-probability-walltime}. It took around 300 hours to model 0.5 minutes and at least 12 minutes needs to be modeled to reach the first experimental data point. However, with an accelerated rate, the computational walltime decreases significantly. Increasing $P_c$ by $10^7$ ($P_c^{acc} = 2.53 \times 10^4$ spikes/nm$^3$s) reduces the computational walltime to model 0.5 minutes to around 10 hours. Increasing it by $10^9$ reduces walltime to around 1 hour ($P_c^{acc} = 2.53 \times 10^6$ spikes/nm$^3$s).

The predicted change in the average grain diameter with time, shown in \cref{fig:Impact-of-probability-grain-size}, decreases slightly with probability $P^{acc}_c \leq 2.53\times10^4$ spikes/nm$^3$ s, possibly due to stochastic changes in the spike locations. However, it increases significantly with larger probabilities. This is because with $P^{acc}_c = 2.53\times10^4$ spikes/nm$^3$s, there is an average of 1.3 thermal spikes per time step. For larger probabilities, there are multiple thermal spikes per time step, and interaction between them raises the temperature and accelerates grain growth. This confirms that using an accelerated rate does not significantly impact the results as long as there are no multiple interacting thermal spikes in each time step. Hence, we use $P^{acc}_c = 2.53\times10^4$ spikes/nm$^3$s in our model and use Eq.~\eqref{eq:P_acc} to calculate the actual time. Note that simulations with this $P^{acc}_c$ are still computationally expensive, requiring 30 hours to simulate just 2 minutes of simulation time, but they are feasible to carry out.



\section{Results}
\label{sec:results}

\begin{figure}[tbp]
\centering
\includegraphics[width=0.99\linewidth]{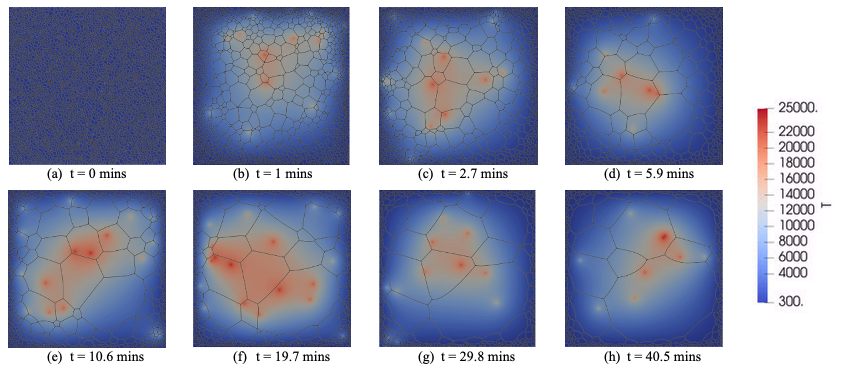}
\caption{Results from 2D simulations of irradiation-assisted grain growth at 300 K assuming a spike duration of $t_s = 100$ ps in a $710\times710$ nm UO$_2$ thin film with 5000 initial grains. Images of the polycrystal domain at different times are shown in (a) - (h). The images are shaded by the temperature in K and the GBs are shown in black.}
\label{fig:Microstructure_Evolution_IAGG}
\end{figure}

We apply our irradiation-assisted grain growth model to predict the evolution in a 710 nm by 710 nm domain with 5000 initial grains that is undergoing irradiation at 300 K. The grain growth is modeled for 40.5 minutes, assuming a thermal spike duration of 100 ps. The simulation takes 969.24 hours (40.38 days) to complete on 160 processors. Images of the polycrystal at various times during the 40.5 minute irradiation are shown in~\cref{fig:Microstructure_Evolution_IAGG}. Due to the presence of thermal spikes, large gradients in temperature occur. The temperature is very high at the center of a spike, up to 25,000 K, but rapidly decreases moving outward from the center. These maximum temperatures are much higher than the melting temperature, but only remain so for a few time steps. The temperatures are higher for thermal spikes near the center of the domain than near the boundary due to the heat flux out of the boundary keeping the temperature at 300 K. These high local temperatures result in localized bursts of grain growth near the thermal spikes. Thus, the grains in the center of the domain grow very quickly, while the grains at the boundaries grow more slowly due to the low temperatures. 

In the initial stages, the grains grow rapidly due to the numerous thermal spikes hitting at or near the GBs. These events result in significant boundary movement at the initial stage of the simulation, such as $t = 1$, 2.7, and 5.9 minutes in~\cref{fig:Microstructure_Evolution_IAGG}. However, at later stages of the simulation, such as $t = 29.8$, 40.5 minutes, fewer GBs are being hit by thermal spike events. Consequently, the growth rate slows in these later stages.

\begin{figure}[tbp]
\centering
\includegraphics[width=0.6\linewidth]{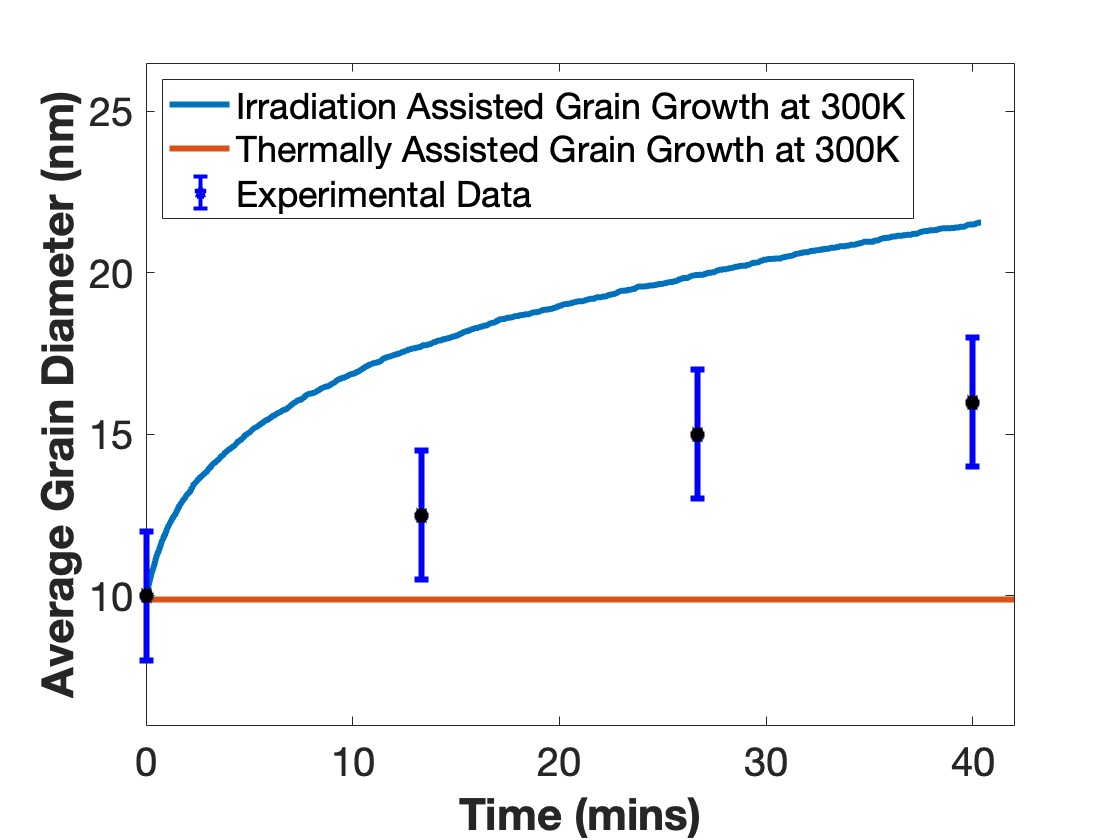}
\caption{Grain diameter versus time with a spike duration of $t_s = 100$ ps at 300 K. Simulation results with no irradiation and the experimental data \cite{Zefeng_2022} are also shown.}
\label{fig:Irradiation_vs_Thermal}
\end{figure}

\Cref{fig:Irradiation_vs_Thermal} shows the average grain diameter with time from our simulation that assumes a spike duration of 100 ps. The average grain diameter quickly increases, especially compared to the results from a simulation at the same temperature but without irradiation, which shows no change in grain size. The growth rate predicted by our simulations with a 100 ps spike duration is significantly higher than was observed in Yu et al.'s experiments~\cite{Zefeng_2022}, even considering the uncertainty in the simulation results. However, as was mentioned in \cref{sec:domain_size_parameter}, we do not know the correct value for the spike duration. Therefore, we use the spike duration value to fit the modeling prediction with the 300 K experimental data.

\begin{figure}[tbp]
\centering
\includegraphics[width=0.6\linewidth]{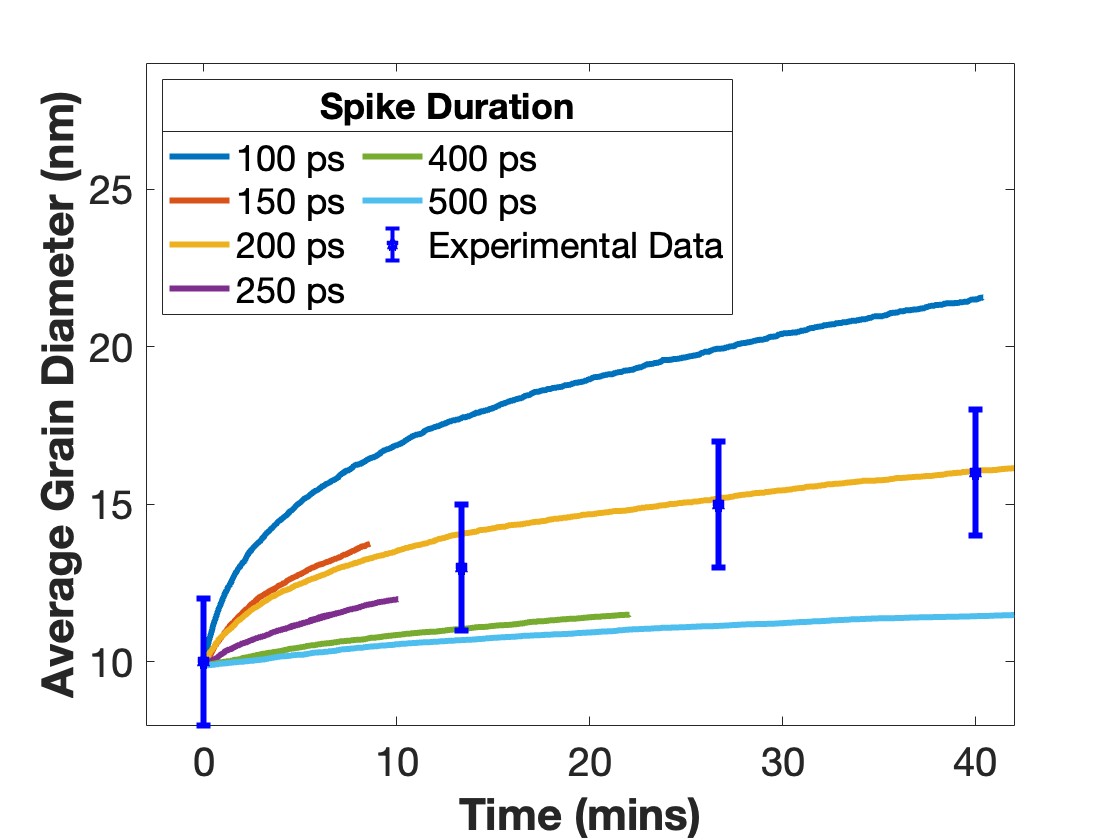}
\caption{Average grain diameter versus time at 300 K for spike durations ranging from 100 ps to 500 ps. The experimental data \cite{Zefeng_2022} are shown for reference.}
\label{fig:Impact_of_spike_duration}
\end{figure}

We carry out simulations with spike durations ranging from 100 to 500 ps. To conserve computational time, we only simulate the full 60 min with the smallest (100 ps) and largest (500 ps) spike durations, and then simulate shorter times for the intermediate values. Note that the computational expense decreases with increasing spike duration, since the minimum time step size is equal to the spike duration. The grain growth decreases significantly with increasing spike duration, as shown in~\cref{fig:Impact_of_spike_duration}. This occurs because the magnitude of the heat generation goes down with increasing spike duration as shown in Eq.~\eqref{eq:heat_gen}. Spike durations ranging from 150 to 250 ps result in growth behavior within the uncertainty ranges of the experiments. A spike duration of 200 ps results in the the best fit with the experimental data.

\begin{figure}[!htbp]
\centering
  \includegraphics[width=0.6\linewidth]{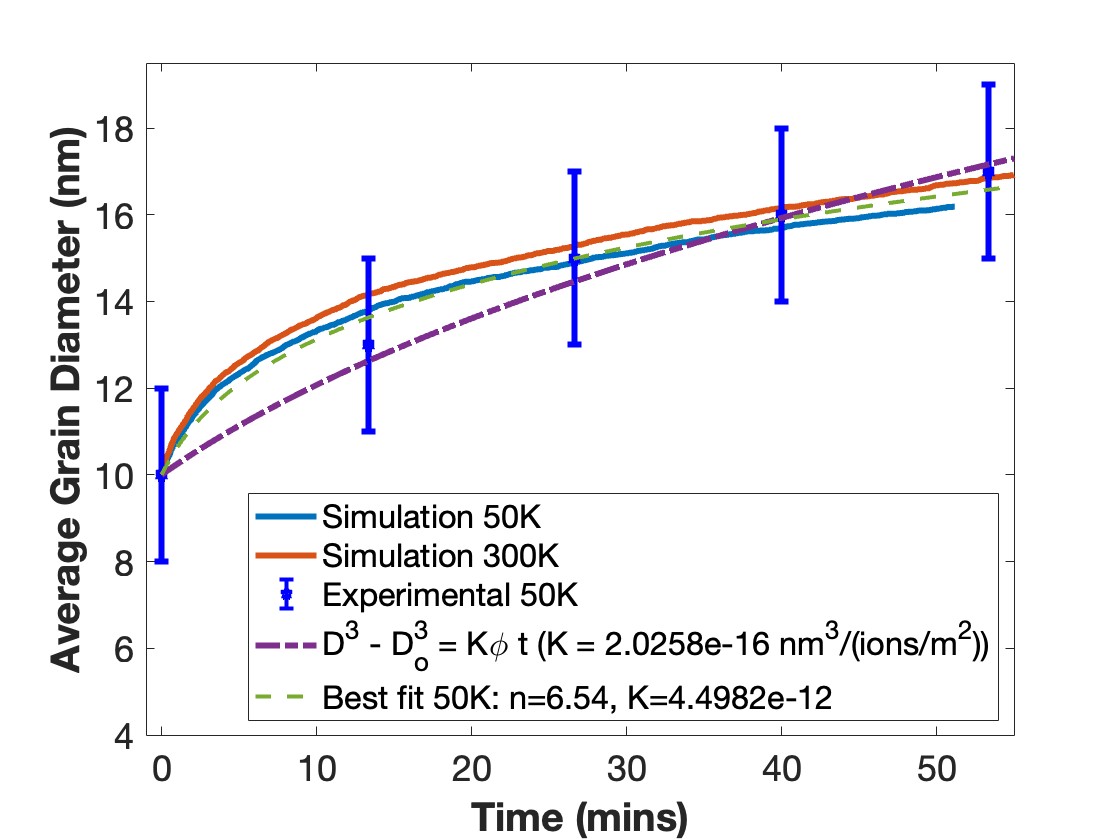}
\caption{Average grain diameter versus time at 50~K using a spike duration of 200~ps. Results at $T = 300$~K are shown for comparison. The experimental data from Yu et al.~\cite{Zefeng_2022} are included for reference. Also shown are the analytical rate-theory prediction based on $D^{n} - D_{0}^{n} = K \phi t$ with $n = 3$, and the best-fit kinetic curve obtained for the 50~K data.}
\label{fig:Validation}
\end{figure}

Having determined a spike duration for which our results compare well with the data at 300 K, we now use it to model the grain growth at 50 K.  The predicted average grain diameters versus time at 50 K are shown in \cref{fig:Validation}, along with the corresponding experimental data. In addition to the simulation results, we include the analytical prediction
based on the rate theory relation $D^{n} - D_{n}^{3} = K \phi t$, where experimental analysis at 50~K has shown that the kinetics are well represented by $n = 3$ \cite{Zefeng_2022}.
We therefore plot the analytical curve corresponding to
$n = 3$, for which we obtain $K = 2.0258\times10^{-16}~
\mathrm{nm^{3}/(ions/m^{2})}$, in close agreement with the value
$K = 9.64\times 10^{-17}~\mathrm{nm^{3}/(ions/m^{2})}$ reported in Ref.~\cite{Zefeng_2022}. To obtain a best-fit curve for the 50~K data, we consider a generalized grain growth law of the form $D^{n} - D_{0}^{n} = K \phi t$, with an initial grain
diameter $D_{0} = 10~\mathrm{nm}$ and a constant ion flux
$\phi = 6.25\times 10^{15}~\mathrm{ions\,m^{-2}\,s^{-1}}$. The parameters $K$ and $n$ are determined by minimizing the sum of squared differences between the model prediction and the experimental average grain diameters at 50~K using the
Nelder--Mead simplex nonlinear least-squares method, which yields $n = 6.54$ and $K = 4.50\times 10^{-12}~
\mathrm{nm}^{n}\,\mathrm{(ions/m^{2})^{-1}}$. This defines the best-fit curve shown in \cref{fig:Validation}. The evolution of the grain diameter with time at 50 K is slightly slower than at 300 K. The temperature at the outer boundary is much lower, but the thermal spike temperatures driving the accelerated grain growth are still very high, resulting in the grain growth rate being nearly athermal. Overall, the simulation results, together with the analytical $n=3$ rate-theory curve, show good agreement with the experimental measurements at 50~K. The consistency among these different representations provides additional validation that the model accurately captures the irradiation-induced grain-growth kinetics at lower temperatures (e.g. 50~K) for nanocrystalline $UO_2$.


\section{Discussion} \label{sec:discussion}

This model represents a large step forward in modeling irradiation-assisted grain growth. However, several assumptions and approximations were made in the simulations and it is important to consider their impact on the predicted behavior. First, we have assumed that the grain structure in the thin film samples is columnar, i.e., it doesn't change through the thickness of the film. This assumption enabled us to model the grain growth in 2D, reducing the computational cost. The initial average grain size of the thin films used in the experiments was around 10 nm and the sample thickness was 50 nm \cite{Zefeng_2022}. Therefore, it is reasonable that multiple grains could exist through the thickness, making our assumption of a columnar grain structure inaccurate. Modeling the grain growth behavior in 3D would provide a more accurate description of the growth behavior, but would also make the simulations very computationally expensive. However, 3D simulations could be carried out in the future to identify any significant differences in the behavior.

A second assumption made in our model is that no grain growth occurs between thermal spikes, so we can skip that time without missing any grain growth. This allows us to accelerate the simulations by artificially increasing the rate of a thermal spike. Our simulations shown in \cref{fig:Irradiation_vs_Thermal} demonstrate that no thermal grain growth occurs at 300 K, lending credence to this assumption. In addition, our results from \cref{fig:Impact-of-probability-walltime} show that as long as we choose a small enough accelerated rate to ensure only one or two spikes occur during a time step, the impact on the average grain diameter with time is small. 

Our final assumption is that the change in the GB mobility with temperature still follows an Arrhenius expression for temperatures above the melting temperature. When a thermal spike occurs within a material, it can experience localized melting that persists for a few picoseconds \cite{de1987role,meldrum1998transient}. Molecular dynamics simulations show that when the thermal spike overlaps a GB, the GB locally loses its character within the thermal spike as the material amorphizes but then reforms, having moved somewhat in that region \cite{kedharnath2019atomistic}. These observations demonstrate that the temporary amorphization results in faster motion of the GB, but do not allow us to quantify how much faster it should be. Our assumption that it continues to follow the same Arrhenius expression is likely inaccurate, but there are no data available in the literature to provide better information about how the effective GB mobility changes in the thermal spike. The calibration of the spike duration to the measured grain growth rates from the experiments helps to compensate for the inaccuracy of this assumption.

Our modeling approach is somewhat similar to that from Bufford et al.~\cite{bufford2015unraveling}. We both use a phase field grain growth model and stochastically introduce thermal spikes into the polycrystalline domain to accelerate the grain growth. However, the approach used to introduce the effect of the thermal spikes varies. They locally increase the GB mobility by an arbitrary amount and qualitatively compare the predicted grain growth to experimental grain growth in gold thin films. We directly represent the heat introduced into the material by the thermal spike and solve the heat equation to dissipate that heat through the material. Our approach allows a closer connection to the irradiation conditions and allows us to do a quantitative comparison with the experimental data, though some calibration is required. Another important difference between the two models is their results. In the model by Bufford et al.\ the irradiation results in somewhat faster grain growth than in thermal conditions.  Our model predicts grain growth in conditions in which no thermal grain growth would normally occur. Thus, our model represents an important step forward in the mechanistic modeling of irradiation-assisted grain growth at the mesoscale.

Our simulations capture the grain growth behaviors shown in the experiments by Yu et al. \cite{Zefeng_2022}. Since our model is based on the hypothesized mechanism that irradiation-assisted grain growth is caused by thermal spikes, this supports this mechanism. This same mechanism is the basis of the mesoscale model developed by Buford et al.~\cite{bufford2015unraveling} and analytical model developed by Kaoumi et al.~\cite{kaoumi2008thermal}. However, this does not eliminate other possible mechanisms for irradiation-assisted grain growth that are simply not dominant under the irradiation conditions used in the experiments by Yu et al.~\cite{Zefeng_2022}. Thus, it would be interesting to investigate the impact of radiation on grain growth in conditions where thermal spikes would not occur. For example, experiments could be carried out using low energy irradiations which would add defects to the system but not cause a thermal spike.

The primary focus of this work is to develop a mechanistic model of irradiation-assisted grain grain in UO$_2$ so we can assess its impact on light water reactor fuel performance. Our simulations show rapid grain growth at first, even at low temperatures of 50 K and 300 K, but that the growth rate quickly slows down as the grain size increases. The average grain size quickly increases from the initial average grain size of 10 nm, but never gets higher than 15 or 20 nm. In the UO$_2$ experiments carried out by Yu et al.~\cite{Zefeng_2022}, the average grain size never increased above 35 nm for all temperatures and irradiation conditions, as shown in \cref{fig:Experimental_Data_Temperatures}. Thus, from our results and from the experiments, it is clear that the radiation enhancement is only significant for small grain sizes. This has also been observed in experiments showing irradiation-assisted grain growth in other materials \cite{KAOUMI_2006490,kaoumi2008thermal}. This occurs because as the grains grow, it is less likely that the thermal spike will overlap a GB, as illustrated in \cref{fig:Microstructure_Evolution_IAGG}.

\begin{figure}[tbp]
    \centering
    \includegraphics[width=0.6\linewidth]{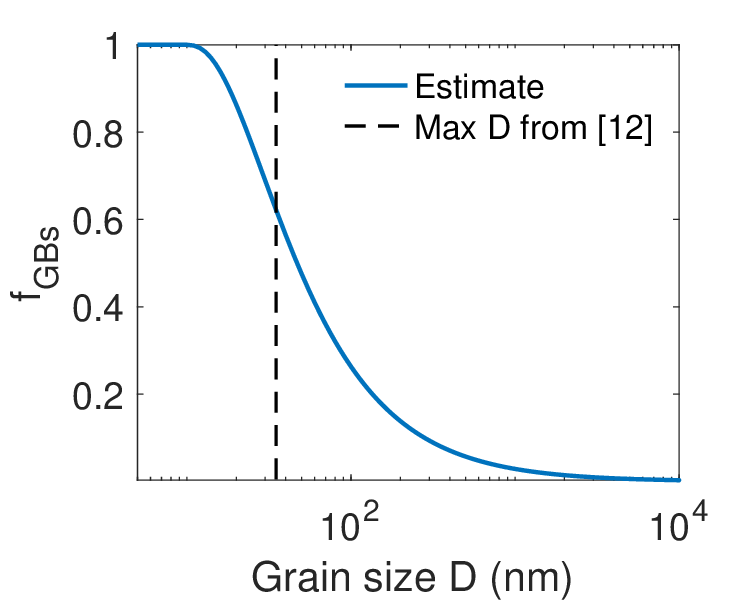}
    \caption{Estimate of the fraction of thermal spikes that touch a GB calculated using \cref{eq:f_estimate}. It is plotted for grain sizes ranging from 5 nm to 10 $\mu$m, assuming a spherical grain. The largest grain size reached in the experiments by Yu et al.~\cite{Zefeng_2022} is shown with a dashed black line.}
    \label{fig:fraction}
\end{figure}
It is valuable to have a more quantitative approach for determining if irradiation-assisted grain growth needs to be considered in a given system. We can estimate this by considering a spherical grain of radius $R$, a thermal spike rate $P_s$, and a spike radius $r_s$. The total number of spikes that occur within this grain on average per unit time can be estimated by multiplying the rate by the grain volume to get $N_T = P_s 4/3 \pi R^3$. In order for a thermal spike occurring within the grain to impact the GB, its center must lie within a distance $\leq r_s$ from the GB. Thus, the average number of spikes per unit time that occur within the grain interior but not touching the GB can be estimated as $N_{int} = P_s 4/3 \pi (R-r_s)^3$. By dividing $N_{int} / N_T$ and subtracting it from one, we can estimate the fraction of spikes that interact with the GB as 
\begin{equation}
    f_{GBs} = 1 - \frac{N_{int}}{N_T} = 1 - (1-r_s/R)^3 \label{eq:f_estimate}
\end{equation}
This estimated fraction for grain sizes $D=2R$ ranging from 5 nm to 10 $\mu$m and $r_s=4.84$ nm (see \cref{table: Experimental and MD data for phase field simulations}) is shown in \cref{fig:fraction}. 

The estimated fraction drops quickly once the grain size is greater than $2r_s$. At 35 nm, the largest average grain size seen in the samples irradiated by Yu et al.~\cite{Zefeng_2022}, the estimated fraction is around 60\%. However, the largest grains will be 1.5 to 2 times larger than the average, with fractions around 30\%. So, the impact of the thermal spikes would be small for the larger grains that are growing. Thus, this estimate agrees reasonably well with the experimental data and our simulation results. The estimate predicts that at grain size of 10 $\mu$m, which is typical for fresh fuel, the fraction of thermal spikes that would contact a GB would be around 0.3\%. Thus, irradiation-assisted grain growth can clearly be neglected for fresh fuel. High burnup structure (HBS) that can form during reactor operation, with an average grain size of hundreds of nm \cite{rondinella2010high}, would have fractions of thermal spikes impacting the GBs around 20\% or lower, and so irradiation-assisted grain growth would also have little impact on HBS. From this, we conclude that irradiation-enhanced grain growth can be neglected when considering the performance of typical UO$_2$ fuel pellets.

\section{Conclusion}
\label{sec:conclusions}
We have developed a mesoscale model of irradiation-assisted grain growth using MARMOT. This model represents the impact of thermal spikes on grain growth by coupling a phase-field grain growth model with a heat conduction equation. Thermal spikes are represented by including a stochastic heat generation term and using the evolving temperature to determine the local grain boundary mobility. All of the parameters needed for the model were taken from the literature, except for the thermal spike duration. The model has been applied to simulate room temperature ion irradiations of UO$_2$ thin films, as documented in the work by Yu et al.~\cite{Zefeng_2022}. Assuming a 100 ps spike duration results in significant grain growth at 300 K, even though thermally activated grain growth would not occur at this temperature. However, the growth rate was higher than that observed in the experiments. By fitting to the 300 K data, we determined that a spike duls
ration of 200 ps provides the best match between simulation results and experiments. Using this spike duration, the results of the grain growth at 50 K compare well with data from Yu et al.~\cite{Zefeng_2022}. These results provide evidence that thermal spikes are the primary mechanism for the irradiation-assisted grain growth observed by Yu et al. Our results also indicate that irradiation-assisted grain growth is only significant for grain sizes smaller than 35 nm; therefore, it does not need to be considered in fuel performance codes.

\section*{Credit Authorship Contribution Statement}
\textbf{Md Ali Muntaha}: Investigation, Methodology, Analysis, Software,
Writing – original draft. \textbf{Larry Aagesen}: Conceptualization, Investigation, Supervision,Writing – review and editing. \textbf{Michael R. Tonks}: Conceptualization, Funding acquisition, Project administration, Supervision, Writing – review and editing.

\section*{Declaration of Competing Interest}
The authors declare that they have no known competing financial interests or personal relationships that could have appeared to influence the work reported in this paper.

\section*{Acknowledgements}
We would like to thank Brandon Battas for valuable review suggestions. We express our gratitude for the high-performance computing resources provided by the University of Florida cluster Hipergator. Also, this research made use of Idaho National Laboratory's High Performance Computing systems located at the Collaborative Computing Center and supported by the Office of Nuclear Energy of the U.S. Department of Energy and the Nuclear Science User Facilities under Contract No. DE-AC07-05ID14517 \cite{Parker_2023, Parker_2024}

This material is based upon work supported by the U. S. Department of Energy (D.O.E), Office of Nuclear Energy University Program Project 17-12797 that was led by Arthur Motta at Pennsylvania State University. We gratefully acknowledge Arthur Motta for valuable discussions and for providing access to the experimental
data used in this work.


\section*{Data Availability}
The MOOSE input files used to generate the simulation results in this paper can be obtained from the authors upon reasonable request.

\bibliographystyle{elsarticle-num.bst}
\bibliography{My_Collection.bib}

\newpage

\end{document}